\documentclass[final,5p,times]{elsarticle}
\usepackage{graphicx}
\usepackage{amssymb}

\journal{Nuclear Instruments and Methods A}

\begin{document}

\begin{frontmatter}

\title{The Tunka-133 EAS Cherenkov light array: status of 2011 }


\author[i1]{S.F.Berezhnev}
\author[i7]{D.Besson}
\author[i2]{N.M.Budnev}
\author[i5]{A.Chiavassa}
\author[i2]{O.A.Chvalaev}
\author[i2]{O.A.Gress}
\author[i2]{A.N.Dyachok}
\author[i1]{S.N.Epimakhov}
\author[i8]{A.Haungs}
\author[i1]{N.I.Karpov}
\author[i1]{N.N.Kalmykov}
\author[i2]{E.N.Konstantinov}
\author[i2]{A.V.Korobchenko} 
\author[i1]{E.E.Korosteleva}
\author[i1]{V.A.Kozhin}
\author[i1]{L.A.Kuzmichev}
\author[i3]{B.K.Lubsandorzhiev}
\author[i3]{N.B.Lubsandorzhiev}
\author[i2]{R.R.Mirgazov}
\author[i1]{M.I.Panasyuk} 
\author[i2]{L.V.Pankov} 
\author[i1]{E.G.Popova}
\author[i1]{V.V.Prosin}
\author[i4]{V.S.Ptuskin} 
\author[i2]{Yu.A.Semeney}
\author[i3]{B.A.Shaibonov(junior)}
\author[i1]{A.A.Silaev}
\author[i1]{A.A.Silaev(junior)}
\author[i1]{A.V.Skurikhin}
\author[i7]{J.Snyder}
\author[i6]{C.Spiering}
\author[i8]{F.G.Schr\"oder}
\author[i7]{M.Stockham}
\author[i1]{L.G.Sveshnikova}
\author[i6]{R.Wischnewski} 
\author[i1]{I.V.Yashin}
\author[i2]{A.V.Zagorodnikov}

\address[i1]{Skobeltsyn Institute of Nuclear Physics MSU, Moscow, Russia}
\address[i2]{Institute of Applied Physics ISU, Irkutsk, Russia}
\address[i3]{Institute for Nuclear Research of RAS, Moscow, Russia}
\address[i4]{IZMIRAN, Troitsk, Moscow Region, Russia}
\address[i5]{Dipartimento di Fisica Generale Universiteta di Torino aDipartimento and INFN,
 Torino, Italy}
\address[i6]{DESY, Zeuthen, Germany}
\address[i7]{Department of Physics and Astronomy, University of Kansas,USA}
\address[i8]{Karlsruhe Institute of Technology, Institut f\"ur Kernphysik, Karlsruhe, Germany}

\begin{abstract}
A new EAS Cherenkov light array, Tunka-133, with $\sim$1\,km$^2$ geometrical
 area has been installed at the Tunka Valley 
(50 km from Lake Baikal) in 2009. The array permits a detailed study of cosmic
ray energy spectrum and  
mass composition in the energy range 10$^{16}$ - 10$^{18}$ eV with a uniform
method. 
 We describe the array construction, DAQ and methods of the array calibration.
 The method of energy reconstruction and absolute  
calibration of measurements are discussed. The analysis of spatial and time
structure of EAS Cherenkov light allows to 
estimate the depth of the EAS maximum $X_{max}$.

The results on the all particles energy spectrum and the mean depth of the EAS
maximum $X_{max}$ vs. primary energy derived from the data of two
winter seasons (2009 -- 2011), are presented.
 Preliminary results of joint operation of the Cherenkov array with antennas for
 detection of 
 EAS radio signals are shown.
Plans for future upgrades -- deployment of remote clusters, radioantennas 
 and a scintillator detector network and a prototype of the HiSCORE
 gamma-telescope -- are discussed. 

\end{abstract}


\end{frontmatter}


\section{Introduction}
\label{s1}


The study of primary energy spectrum and mass composition in the energy range
$10^{15}$ - $10^{18}$ eV is of crucial importance for the understanding of the
cosmic rays origin and propagation in the Galaxy. 

To measure the primary cosmic ray energy
spectrum and mass composition in the mentioned energy range,  
the new array Tunka-133 (\cite{1}, \cite{2}), with nearly 1 km$^2$ geometrical
area has been deployed in the Tunka Valley, Siberia. It
records EAS Cherenkov light using the atmosphere of the Earth as a huge
calorimeter and has much better energy resolution ($\sim$ 15\%) than EAS arrays
 detecting only charged particles.

\section{Tunka-133}
\label{s2}

 The Tunka-133 array consists of 133 wide-angle optical detectors  based on the
 PMT EMI 9350 with a hemispherical photocathode of \mbox{20\,cm} diameter.
The detectors are grouped into 19 clusters, each cluster with seven detectors --
six 
hexagonally arranged detectors and one in the center. The distance between the
detectors inside the cluster is 85 m. 


The Cherenkov light pulse is sent via 95 m coaxial cable RG58 to the center of
a cluster and digitized. 
 The dynamic range of the
amplitude measurement is about $3\cdot 10^4$. This is achieved by means of
two channels for each detector extracting the signals from the anode and from an
intermediate dynode of the PMT with different additional amplification factors.

\begin{figure}[!t]
\vspace{3mm}
\centering
\includegraphics[width=0.9\columnwidth]{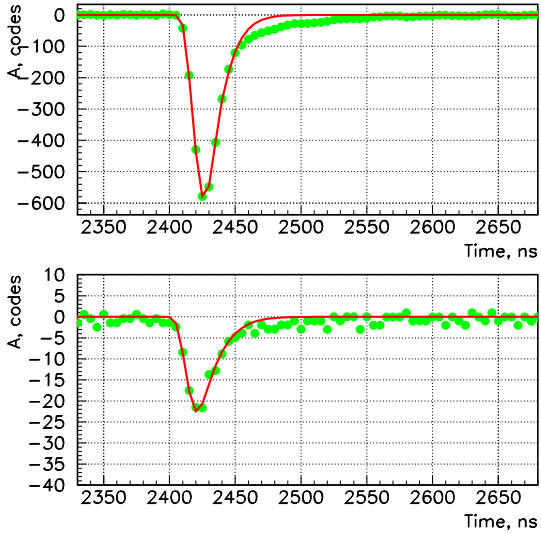}
\caption{: Example of a pulse from one Tunka-133 detector, recorded by high and low gain channels}
\label{fig1}
\end{figure}

 The cluster electronics includes the cluster controller, 4 four-channel
FADC boards, an adapter unit for connection with optical dtectors and a special
temperature controller. The 12 bit and 200 MHz sampling FADC board is based on
AD9430 fast ADCs and  
FPGA XILINX Spartan XC3S300 microchips. The cluster controller consists of an
optical 
transceiver, a synchonization module, a local time clock and a trigger module.
The optical transceiver operating at 1000 MHz is responsible for data
transmission and 
formation of 100 MHz synchronization signals for the cluster clocks. The cluster
trigger (the local trigger) is formed by the coincidence of at least three
pulses from the 
optical detectors exceeding the threshold within a time window of 0.5$\mu$s. 
The time mark of the local trigger
is fixed by the cluster clock. The accuracy of the time synchronization between
different clusters is about 10\,ns.
Each cluster electonics is connected with the DAQ center by a
multi-wire cable containing four copper wires and four optical fibers.

The central DAQ station consists of 4 DAQ boards synchronized by a single 100
MHz generator. The boards are connected with the master PC by 100 MHz Ethernet
lines.

\section{Data processing and reconstruction of the EAS parameters}
\label{s3}
\subsection{Data processing}

The primary data record for each Cherenkov light
detector contains 1024 amplitude values with step of 5 ns (Fig.\ref{fig1}).
Thus the each pulse waveform is
recorded together with the preceding noise as a trace of 5\,$\mu s$.
To derive the three main parameters of the pulse: front 
delay at a level 0.25 of the maximum amplitude 
$t_i$, 
pulse  area $Q_i$ and
full width at half-maximum (FWHM) $\tau_i$, each pulse is fitted with a
special self-designed smoothing curve \cite{icrc0492}. 

The reconstruction of the EAS core position is performed with two methods -- 
by 
fitting the  measured charges $Q_i$ with the 
lateral distribution function (LDF (\cite{cris_08}) and by
fitting the measured pulse widths 
$\tau_i$ by the width-distance function (WDF)(\cite{cris_2010}).

\subsection{Energy reconstruction}
As a measure of energy we used the Cherenkov light flux density at a core
distance of 175\,m - $Q(175)$. Connection between the EAS energy E$_0$ and 
Q$_{175}$ may be expressed by the following formula:

\begin{equation}
E_0=C\cdot {Q}_{175}^{g}    
	\end{equation} 
 
It 
was found from CORSIKA simulation, that for the energy range of \mbox{$10^{16} -
10^{17}$\,eV} and zenith angle range 
\mbox{$0^{\circ} - 45^{\circ}$} the value of index $g$ is 0.95 for pure
proton composition and 0.91 for pure iron compositon. 
 For energy reconstruction the value of $g$ equal to 0.93 was chosen.

\begin{figure}[!t]
  \vspace*{-3mm}
  \centering
  \includegraphics[width=2.5in]{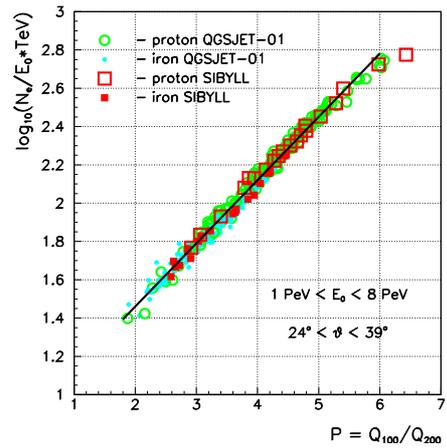}
  \vspace*{-3mm}
  \caption{$N_e/E_0$ vs $P$ }
  \label{fig2}
 \end{figure}

%
%
%
%
%

To reconstruct the EAS energy from Cherenkov light flux one needs to know an 
absolute sensitivity of Cherenkov detectors and the atmosphere transparency. To
avoid these problems, the method of normalization of the integral energy
spectrum to the reference one is used.
The reference energy spectrum was obtained at the QUEST 
 experiment
 \cite{cris_06}. 
In that experiment, five wide-angle Cherenkov 
light detectors were installed at Gran-Sasso for common operation with the
EAS-TOP array.
The analysis of CORSIKA simulations shows a strict correlation between the
size/energy ($N_e/E_0$) ratio and the steepness ($P$) of the EAS Cherenkov light
lateral distribution.

The relation between $N_e/E_0$ and $P$ (Fig.\ref{fig2}) is independent both
from the mass of 
primary particle and the hadronic interaction model used for the simulation and
provides the primary energy from the measurement of $N_e$ and $P$. 
To reconstruct
the LDF steepness $P$, the knowlege of 
PMT absolute sensitivity and the atmosphere transparency is not needed.  The
integral energy spectrum of 
cosmic rays obtained in the QUEST experiment is used as the reference one now.
The integral 
energy spectra obtained for each night of Tunka-133 operation is normalised to
that reference one.

\subsection{Reconstruction of $X_{max}$}

Recording the pulse waveform for each detector allows to use
 two methods of $X_{max}$ reconstruction,
which were developed for our experiment.
The first is based on the shape of the LDF and called $P$-method. 
The second method,
 the $W$-method, is based on an analysis of time width of the Cherenkov pulses.

The LDF
shape is described by an expression with a single parameter, the steepness $P$
\cite{5}. 
$P$ is strictly connected with the distance from the core position
to the EAS maximum \cite{p_hmax}:

\begin{equation}
H_{max}=A- B\cdot P
\end{equation}    

MC simulations shows, that this relation does not depend on energy, zenith
angle of the showers, mass composition and the model of nuclear interaction used
for the simulation.
To get a uniform estimation for $P$  over a wide range of
energies, we remove from the analysis the detectors at core distances more than
200\,m during the last step of parameter reconstruction.

The $W$-method uses the sensitivity of the pulse width 
at some fixed core distance to the position of the EAS maximum. We fixed this
distance to 400\,m and 
recalculated the values measured with detectors between 200 and 400 m away from
the core to this distance. 
To characterize the pulse duration, we use the effective pulse with $\tau$ 

\begin{equation}
\tau = S/(1.24\cdot A_{max}),
\end{equation}
where $S$ is the area under the pulse in $[dig.counts*ns]$, $A_{max}$ in $[dig.counts]$
 is its maximum amplitude.\\
To recalculate the pulse width to 400\,m, the width-distance function (WDF) is
used. This function was constructed on the basis of CORSIKA simulation and
descibed in \cite{cris_2010}. 
It was also shown \cite{cris_2010}, that the value of
$\tau(400)$ is connected with the thickness of the atmosphere between the
detector and $X_{max}$ ($\Delta X_{max} = X_0/cos\theta - X_{max}$) by the
expression:

\begin{equation}
\Delta X_{max} = C - D\cdot log~\tau_{400}.
\end{equation}    
This relation is correct for any primary nucleus, any energy and zenith angle of
the shower and any interaction model, as in the case of LDF steepness mentioned
above.

\section{Energy spectrum}
\label{s4} 
\subsection{Energy spectrum for inside events}

The data taking by the Tunka-133 array started in October 2009 and has been 
continued during two winter seasons of 2009-2010 and 2010-2011. As a result, the
data were collected for 597\,hrs of 102 clean moonless nights. Only the
nights with more than 2\,hr of clean weather were taken into account. The
average trigger rate was about 2\,Hz. The number of recorded events was about $4\cdot 10^{6}$.

\begin{figure}[!h]
  \vspace*{-3mm}
  \centering
  \includegraphics[width=2.9in]{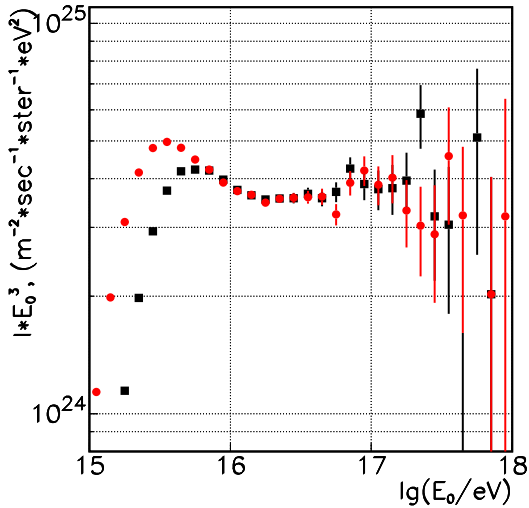}
  \vspace*{-3mm}
  \caption{Energy spectrum for two zenith angle intervals. circles -
  $0^{\circ}< \theta <30^{\circ}$, 
   squares - $30^{\circ}< \theta < 45^{\circ}$}
  \label{fig-tet}
 \end{figure}

\begin{figure}[!h]
  \vspace*{-6mm}
  \centering
  \includegraphics[width=2.9in]{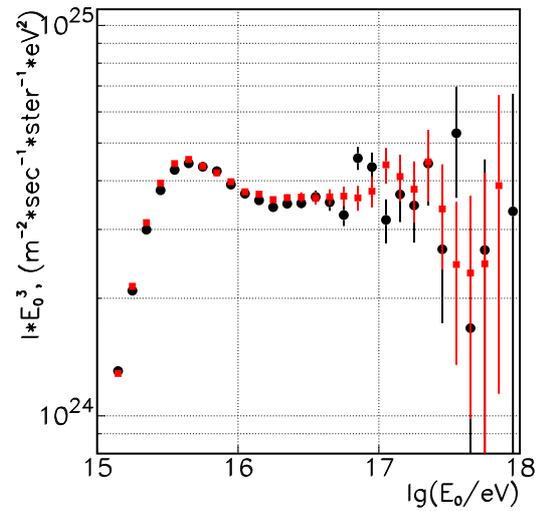}
  \vspace*{-3mm}
  \caption{Different energy spectra for two seasons of the array operation.
  Circles - 2009-2010, squares - 2010-2011} 
  \label{fig4}
 \end{figure}

For construction of the energy spectrum the events 
 with core position inside a circle of radius $R < 450$\,m from the center of the array
 were selected.  The EAS zenith angle for the events collected into the spectrum is limited by
$45^{\circ}$. 
 The threshold energy of 100\% registration efficienty for chosen area and
 zenith angles is $6\cdot10^{15}$\,eV.
 The energy spectra obtained for two different zenith angular ranges ($0^{\circ}
 - 30^{\circ}$ and $30^{\circ} - 45^{\circ}$) 
are in good agreement for energy more than \mbox{$6\cdot10^{15}$\,eV} (Fig.\ref{fig-tet}).
It was found, that the energy spectrum obtained 
for the  first season of the array operation (2009-2010) is in 
good agreement with the energy spectrum from the second season (2010-2011)
(Fig.\ref{fig4}) up to the energy \mbox{$6\cdot 10^{16}$\,eV}. 
The difference in intensity at higher energies seems to be only due to
statistical fluctuation. 

\begin{figure}[!t]
  \vspace*{-3mm}
  \centering
  \includegraphics[width=2.9in]{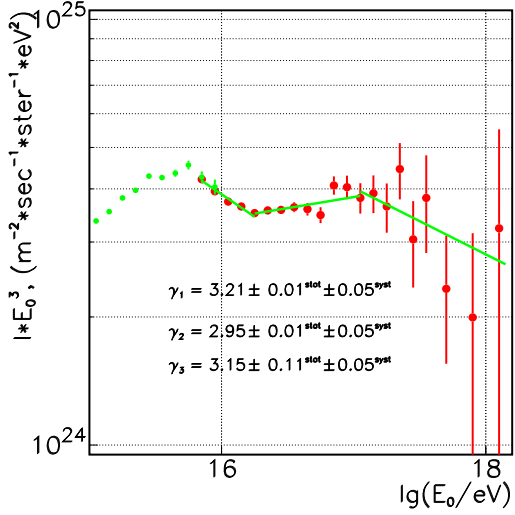}
  \vspace*{-3mm}
  \caption{Summary energy spectrum for two seasons of Tunka-133 operation ( red circles)
   and Tunka-25  
  energy spectrum ( green circles).}
  \label{fig5}
 \end{figure}

The indication of a "bump", found in the first season 
 at the energy \mbox{$8\cdot 10^{16}$\,eV} 
 \cite {Texas}, is not seen for the second season spectrum and 
practically disappears in the combined spectrum (Fig.\ref{fig5}).

 The combined energy spectrum for two seasons of the array operation is presented
 in Fig.\ref{fig5}.  
 It contains $\sim$40000 events
 with energy \mbox{$E_0 > 10^{16}$\,eV} and and $\sim$400 events with \mbox{$E_0
 > 10^{17}$\,eV}. The energy spectrum of Tunka-133 is compared with that of
 Tunka-25 \cite{cris_06}, 
 the predecessor of Tunka-133, in this figure. The energy spectrum beyond the
 knee looks rather complicated.  
  One can see that the  spectrum can be fitted  by  power laws with 3 different
  power law indexes:\\ 
  $3.21\pm(0.01)^{stat}\pm(0.05)^{syst}$ for $6\cdot 10^{15} - 2\cdot
  10^{16}$\,eV,\\
  $2.95\pm(0.01)^{stat}\pm(0.05)^{syst}$ for $2\cdot 10^{16} - 10^{17}$\,eV,\\  
  $3.15\pm(0.11)^{stat}\pm(0.05)^{syst}$ for $10^{17} - 10^{18}$\,eV.\\


\subsection{Energy spectrum for outside events}

\begin{figure}[!h]
  \vspace*{-3mm}
  \centering
  \includegraphics[width=2.9in]{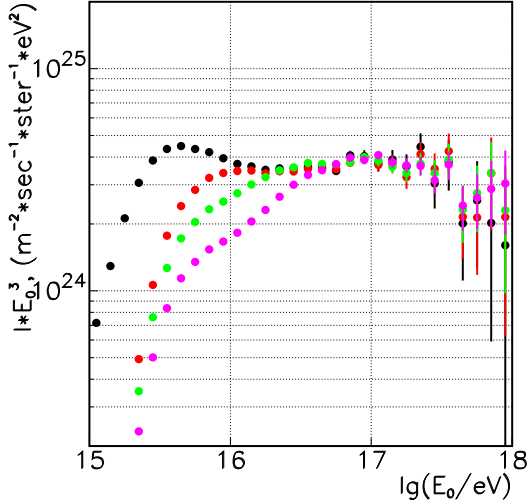}
  \vspace*{-3mm}
  \caption{Comparision of energy spectrum for inside events (black circles) and 
  spectra for outside events $R<550$ m ( red circles), $R<650$ m (green circles) and
  $R<800$ m (violet circles).} 
  \label{fig6}
 \end{figure}

 It was found that the energy
 spectrum including the outside events can be 
 reconstructed too and this spectrum is in good agreement with 
the energy spectrum for inside events beyond some
energy threshold. Fig.\ref{fig6} presents a comparison of the spectrum for 
inside events ($R < 450$\,m) with 
 spectra for events with core inside circles $R<550$\,m, $R<650$\,m and
 $R<800$\,m.     
 
\begin{figure}[!h]
  \vspace*{-6mm}
  \centering
  \includegraphics[width=2.9in]{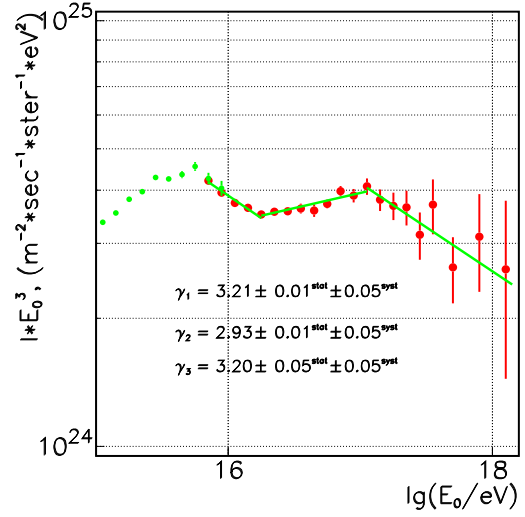}
  \vspace*{-3mm}
  \caption{Combined energy spectrum. Green circles - Tunka-25 data.}
  \label{fig7}
 \end{figure}

It is seen that the threshold energy of 100\% registration efficiency increases
with increasing of the radius $R$, but for  
$E_0 > 6\cdot 10^{16}$\,eV all spectra are in good agreement.
 Based on such the result we construct the combined energy spectrum
 (Fig.\ref{fig7}) from the events with $R<450$\,m 
for  \mbox{$E_0 < 6\cdot 10^{16}$\,eV}
 and events with $R<800$\,m for the 
higher energy. 
This combined  spectrum
 contains about 1200 events with \mbox{$E_0 > 10^{17}$\,eV}.

 The shape of the combined spectrum is nearly the same as for inside events spectrum 
 and can be fitted by the power laws with 3 different  indexes:\\
$3.21\pm(0.01)^{stat}\pm(0.05)^{syst}$ for $6\cdot 10^{15} - 2\cdot 10^{16}$\,eV,\\
$2.93\pm(0.01)^{stat}\pm(0.05)^{syst}$ for $2\cdot 10^{16} - 10^{17}$\,eV,\\  
$3.20\pm(0.06)^{stat}\pm(0.05)^{syst}$ for $10^{17} - 10^{18}$\,eV.\\

\subsection{Comparison with results of other experiments}

 \begin{figure}[!b]
  \vspace*{-3mm}
  \centering
  \includegraphics[width=2.9in]{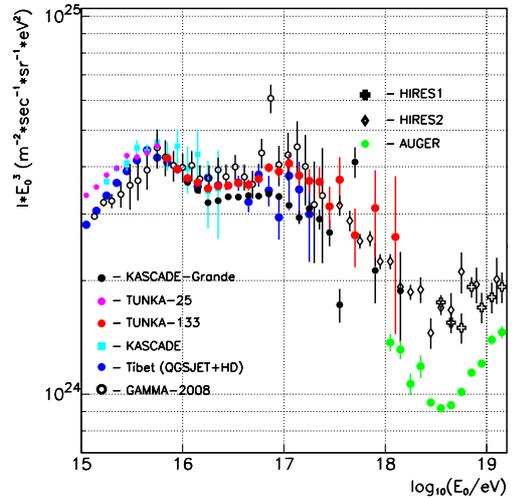}
  \vspace*{-3mm}
  \caption{Tunka-133 all particle energy spectrum in comparison with results of other experiments.}
  \label{fig8}
 \end{figure}

A comparison of the Tunka-133 spectrum with the results of other experiments is
 presented in Fig.\ref{fig8}. 
 Our spectrum is in good agreement 
with spectra of KASCADE(\cite{KASCADE}), 
  Tibet (\cite{Tibet}) and GAMMA (\cite{GAMMA}) arrays.
  The energy range covered by our spectrum (\mbox{$10^{16} - 10^{18}$\,eV}) is
  nearly the same as covered by the 
   KASCADE-Grande array data \cite{K-Gr}.

Both spectra reproduce the same structures: 
decrease of power law index  at $2\cdot 10^{16}$\,eV and 
an increase at $10^{17}$\,eV. 
  The difference in absolute   cosmic ray
flux intensity for Tunka-133 and KASCADE-Grande
  spectra is smaller than 30\%  even at  $10^{17}$\,eV where the difference
  reaches the maximum value. 
  It should be mentioned, that if for $g$ (see expession 1)
  a value of 0.91 is used, the difference 
   in the intensity becomes
  smaller than 10\%.

%

\section{Phenomenological approach}

The parameters derived from CORSIKA simulation may slightly differ from the
experimentally  measured parameters. For instance, the linear relation between  $P$ and
$H_{max}$ observed for MC simulation (equation 2) may hold also for the experiment, but with a
slightly different slope. For our recalculation procedure we used the slope
derived from experimental data, not from MC, i.e choosing a "phenomenological
approach".

\begin{figure}[!b]
\centering
  \includegraphics[width=2.9in]{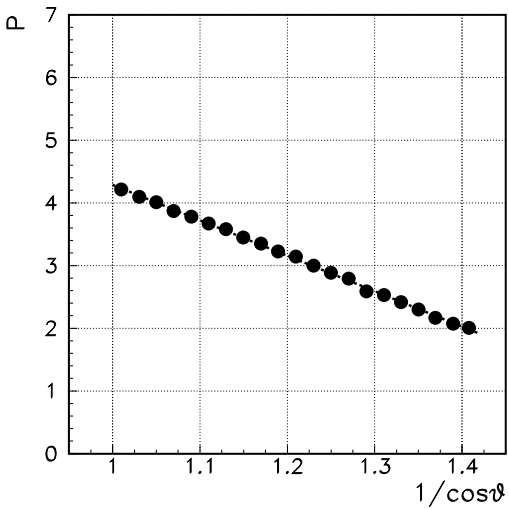}
\caption{ $P$ vs.$\theta$ dependence from experiment.}
\label{fig9}
\end{figure}

In our case we can use the zenith angle dependence of the parameter $P$.
 This experimental dependence for the fixed energy $E_{0} = 10^{16}$~eV is shown
 in Fig.\ref{fig9}.
This mean dependence was constructed using all the 14400 events from the
energy bin $16.0<log_{10}(E_{0}/eV)<16.1$.
The mean zenith angle can be recalculated to the mean distance to the EAS
maximum. To make this recalculation we use the model of the atmosphere from
\cite{atm} for the real experimental 
conditions $<t> = -30^\circ $C and $X_{0} = 965 g\cdot cm^{-2}$. This model
gives the following expression for the inclined distance to the EAS maximum in
km:

\begin{equation}
H_{max}=\frac{H_0}{cos\theta}\cdot 
\Bigl(1-\Bigl(\frac{X_{max}\cdot cos\theta}{X_{0}}\Bigr)^{0.0739}\Bigr)
\end{equation}
where $H_0 = 96~ km$.\\
To fix the absolute value of $<H_{max}>$ we need to fix the
mean value $<X_{max}>$ for the above mentioned energy. The most reliable
experimental estimation of the mean depth of maximum by the data of our previous
Tunka-25 experiment is $<X_{max}> = 560 g\cdot cm^{-2}$ \cite{p_hmax}. 
Recalculated to the mean $<lnA>$, this value is in good agreement with the
results of some other experiments \cite{KASCADE-1}.
The so derived experimental recalculated dependence $H_{max}$ vs. $P$ is
shown in Fig.\ref{fig10}. It can be fitted with a linear expression:

\begin{figure}[!b]
\centering
  \includegraphics[width=2.9in]{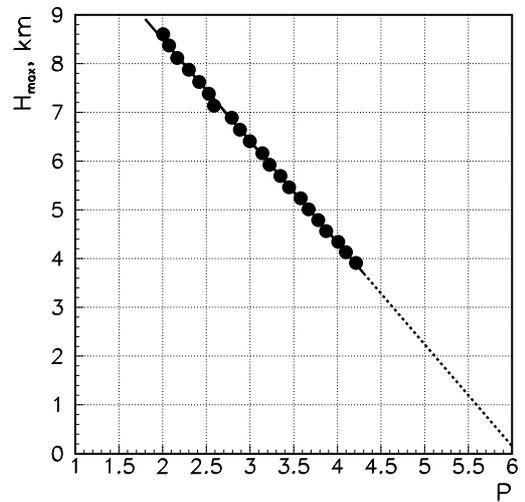}
\caption{  $H_{max}$ vs.$P$ dependence from experiment.}
\label{fig10}
\end{figure}

\begin{figure}[!t]
\centering
  \includegraphics[width=2.9in]{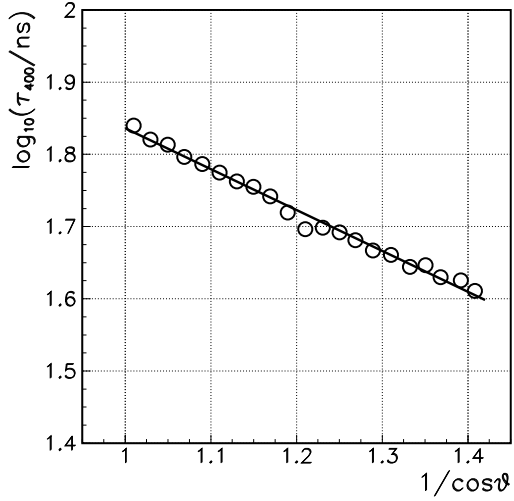}
\caption{ $\tau_{400}$ vs.$\theta$ dependence from experiment.}
\label{fig11}
\end{figure}

\begin{equation}
H_{max}=A_1- B_1\cdot P
\end{equation}    
where $A_1 = 12.67~ km$ and $B_1 = 2.09~ km$.\\
This expression was used for the estimation of $X_{max}$ for each individual
event. 

\begin{figure}[!t]
\centering
  \includegraphics[width=2.9in]{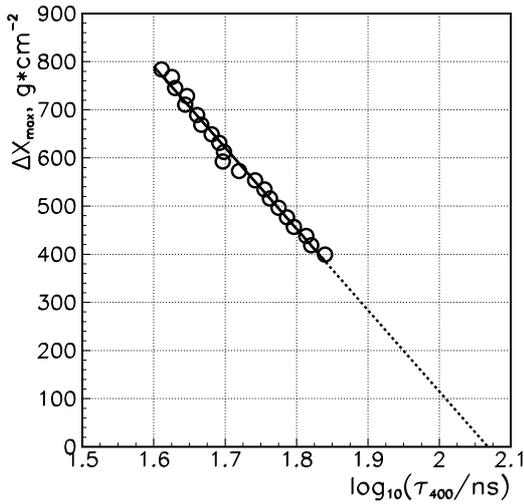}
\caption{  $\Delta X_{max}$ vs.$\tau (400)$ dependence from experiment.}
\label{fig12}
\end{figure}

A similar approach for the second parameter sensitive to the position of EAS
maximum $\tau(400)$ can be used but for higher energy because of the greater
energy threshold of this method. The zenith angular dependence of $\tau_{400}$
fot the logaritmic energy bin $16.5<log_{10}(E_{0}/eV)<16.6$\,eV is shown in
Fig.\ref{fig11}. The second method normalized to the $X_{max}$ obtained by the
first one for the energy
$3\cdot 10^{16}$\,eV to fix the absolute scale of $\Delta X_{max}$.  This
approach results in an expression connecting $\tau_{400}$ 
with the thickness of 
matter between the array and the EAS maximum in $g\cdot cm^{-2}$:

\begin{equation}
	\Delta X_{max}=C_1- D_1\cdot log~ \tau(400)
\end{equation}		
where $C_1 = 3493~ g\cdot cm^{-2}$, $D_1 = 1689~ g\cdot cm^{-2}$.\\
The result is shown in Fig.\ref{fig12}.


\section{Experiment: $<X_{max}>$ vs. $E_{0}$}
      
\begin{figure}[!t]
\centering
  \includegraphics[width=2.9in]{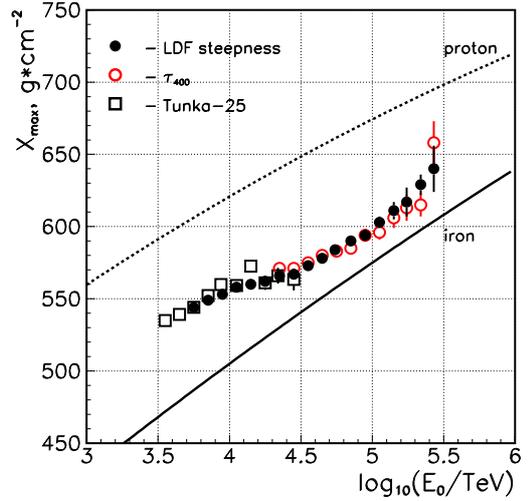}
\caption{ Experimental $<X_{max}>$ vs.$E_0$ dependence for Tunka-133,
compared with Tunka-25.}
\label{fig13}
\end{figure}

The experimental dependence of mean $<X_{max}>$ vs. primary energy $E_{0}$
obtained with two methods described above in the energy range $5\cdot 10^{15} -
3\cdot 10^{17}$\,eV is presented in Fig.\ref{fig13}. The new measurements are
compared with 
that obtained with our previous array Tunka-25 and with the theoretical curves
simulated with QGSJET-01 model for primary protons and iron nuclea. 
The first conclusion is, that the threshold of the  $W$-method is higher
than that of the LDF steepness method, but the experimental points obtained by
both methods coincide within the statistical errors.

Much higher
statistics of Tunka-133 points has led to the much smoother behavior of the
experimental dependence 
compared with the Tunka-25 data. The experimental
points go closer to the iron curve with energy grow from the knee to about
$10^{17}$\,eV. 
There is a tendency of a backwards movement of the experimental
points to the proton curve at the energy more than 
$10^{17}$\,eV, but the
statistical errors are too big to insist on such conclusion.  

\begin{figure}[!b]
\centering
  \includegraphics[width=2.9in]{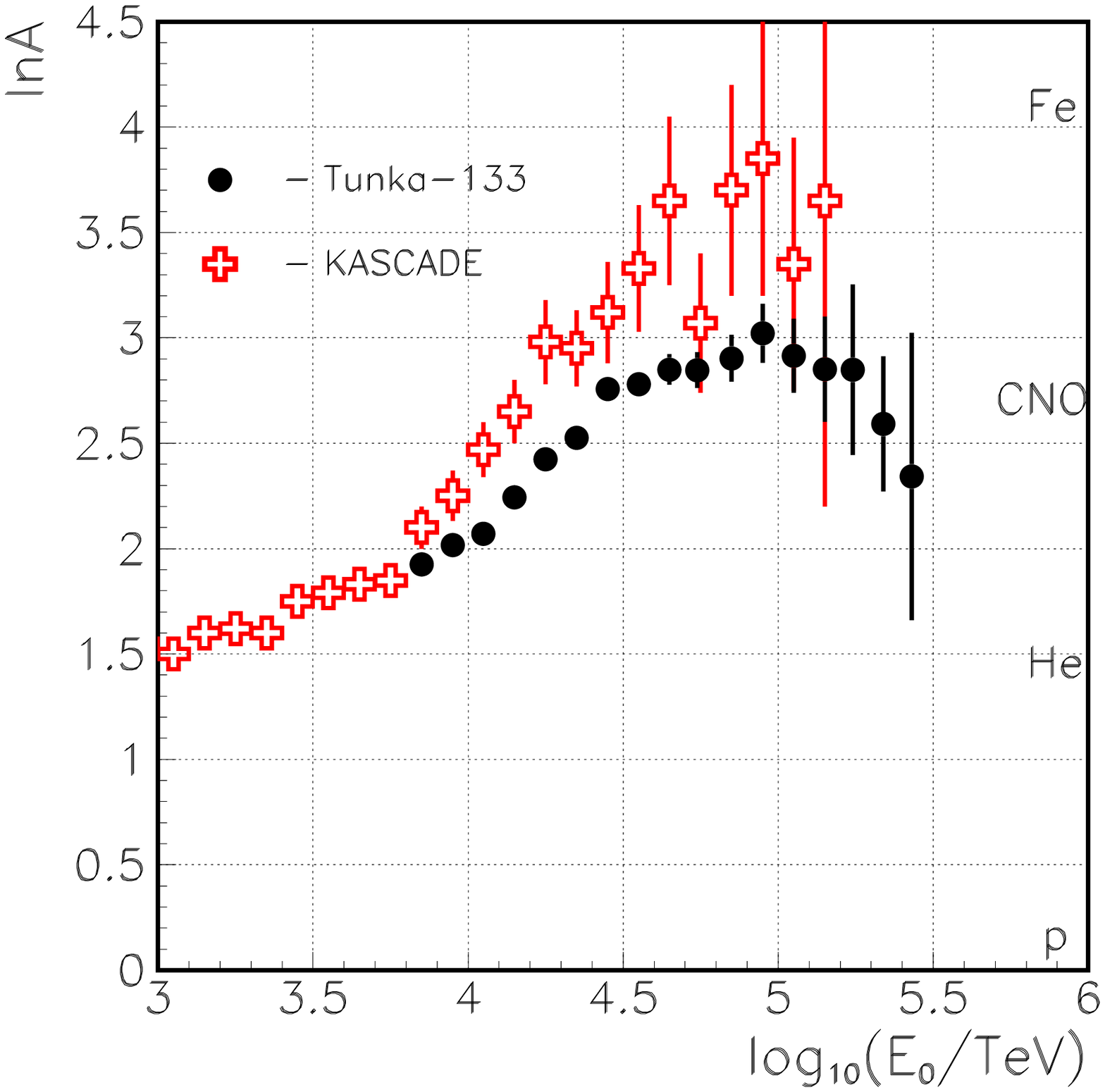}
\caption{Experimental $<lnA>$ vs.$E_0$ dependence.}
\label{fig14}
\end{figure}

The mean values of $<X_{max}>$ can be recalculated to the mean values of $<lnA>$ 
by a simple method of interpolation taking into account the corrections to the
asymmetry of the $X_{max}$ distribution, estimated 
at our previous work
\cite{cris_08}. The result of 
such approach for the points derived from the LDF
steepness analysis are shown in Fig.\ref{fig14}. 
We have to note that this procedure can
give different absolute values of $<lnA>$ for different supposed model of
nuclear interaction. The model QGSJET-01 we use for the analysis provides the
highest position of the EAS maximum as compared with the
other models used now
for simulations. The most deep position of EAS maximum can be obtained using the
QGSJET-II-03 model. The mean difference in $X_{max}$ between these models is
about 20 $g\cdot cm^{-2}$. Using of this last model can increase the estimation
of $<lnA>$ to about 0.8 for the same experimental value of $<X_{max}>$.


The experimental points are compared with that obtained from the analysis of the
muon/electron ratio in the KASCADE \cite{KASCADE-1} experiment in Fig.\ref{fig14}.

\section{Other Results}
\label{s5}

 \subsection{Measurements of radio signals from EAS}

\begin{figure}[!b]
\centering
  \includegraphics[width=\columnwidth]{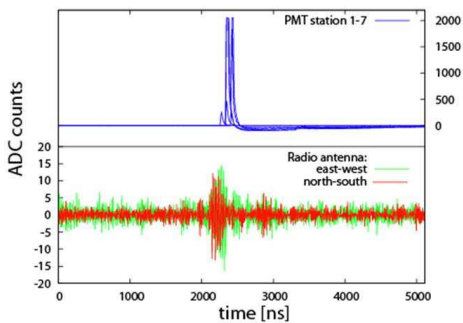}
\caption{One of the radio candidate events. 
A radio pulse is detected in both polarizations shortly before 
the signal of the PMTs.}
\label{fig15}
\end{figure}

To study whether, in addition to the Cherenkov light, also the radio 
emission of air showers can be detected at Tunka, a SALLA
antenna \cite{Kroemer09} was deployed at Tunka cluster 7 in summer 2009.
SALLA is read out simultaneously with the PMTs of that cluster.
Indeed numerous radio pulses could be detected, but most of them are due
to RFI by the Tunka electronics or other background. However,
for Tunka events with a high PMT signal, i.e. a high primary energy,
an accumulation of events with a radio pulse shortly before the PMT
pulses is observed (see Fig.\ref{fig15} for an example). 
Selecting these events, about 70 radio candidate events have been found
within the season 2009/2010.

\begin{figure}[!t]
\centering
  \includegraphics[width=0.9\columnwidth]{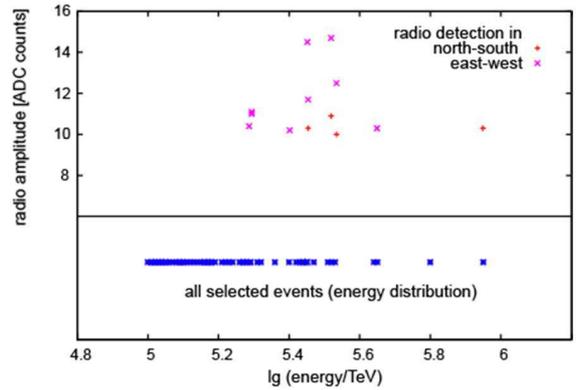}
\caption{Energy distribution of a selection of Tunka events and the radio amplitude for those 
of these events which show a radio pulse in at least one 
polarization: radio pulses are preferred visible for higher energies}
\label{fig16}	
\end{figure}

Several tests have been performed to confirm whether the detected
radio pulses are really from air showers or due to any kind of background 
generated by Tunka.
One test was to check for which events in a selection of high quality
and high energy Tunka events a radio pulse could be detected 
(see Fig.\ref{fig16}).
Consistent with the results of other radio experiments
(e.g., LOPES \cite{Falcke05}), a detection is more likely in the
east-west aligned antenna and for high energies. Furthermore, 
the ratio between the amplitude in the east-west and north-south 
aligned antenna has been compared to REAS3 \cite{Ludwig11} simulations. 
For 7 out of 11 events which show a radio pulse in both polarizations,
the amplitude ratio is compatible within the measurement uncertainties 
estimated with a formula derived for LOPES \cite{Schroeder11}.
This is another indication, that at least most of the detected 
radio candidates are due to the radio emission of air showers.

\subsection{Time calibration}

For time synchronization of the optical modules of the Tunka-133 array, 
a dedicated calibration system has been developed \cite{LED}. The
calibration system is based on a powerful nanosend LED light source especially designed for the
system. The light source consists of four high InGaN/GaN royal blue ($\lambda_{max}$ = 450 nm)
LEDs with 100$^{\circ}$ full emission angle. 
Thus, the light source covers 2$\pi$ azimuth angle.
 
  Each LED of the source is driven by its own avalanche transistor driver and provides $\sim$
  10$^{12}$ photons per pulse, with $\sim$4\,ns pulse width (FWHM). All LED drivers of
  the source are triggered simultaneously. Since the light source is installed a few meters
  above the ground it can, in principle, illuminate the whole array, given that
  all optical modules of the  array are equipped with appropriate reflectors.

\section{Plans for upgrading}

The Tunka-133 facility will be significantly upgraded until fall 2013. In 2011 
we will install additional clusters of optical detectors at large distance 
from the center of the array ("distant clusters"). In 2011-2013, we plan to install radio antennas and
scintillation detectors for common operation with Tunka-133. 

\subsection{Deployment of six distant clusters.}

 To improve the accuracy of energy
reconstruction for the events with the cores located outside the array
geomentry we plan to install six new clusters  
at 1 km radius around the center of Tunka-133 \cite{Texas}. These additional 42
optical detectors will increase the 
effective area for $E_0 > 10^{17}$ eV by a factor of 4. The first distant
cluster was deployed in 
autumn 2010, the next five will be deployed during summer-autumn 2011.

\subsection{ Net of radio anntenas at Tunka}
 
 The positive results of common operation of the first SALLA antenna and the
 Tunka-133 array suggest that it would
 be interesting to install a  net of the same type of antennas at Tunka. The principle aim of such
 a radio-array is to show whether radio measurements of cosmic ray air showers allow 
 the same precision for cosmic ray measurements as Cherenkov light measurements.
 In particular, the precision of
 the energy and mass determination of primary cosmic-ray particles will be investigated. The determination
 of the precision is a crucial input for the design of the next-geneneration large cosmic-ray
 observatories, since radio measurements are not limited to dark, moonless nights as other calorimetric 
 detection techniques, like Cherenkov or fluorescence light measurements.
 
  In autumn 2011 we plan to install additional 2-3 antennas, further $\sim$20 antennas will be 
installed in 2012.

\subsection{Scintillation detectors}

 The deployment of scintillation counters within the Tunka array provides a 
 cross-calibration of different methods of air shower parameters reconstruction.
   
 In 2010, a test variant of a scintillation detector was installed. It can
 operate independently as well as triggered by Tunka-133. The mass production of
 detectors will start at the beginning of 2012. We plan to produce 20 scintillation detectors
 with 10\,m$^2$ area each. 
  As alternative possibility, it is under discussion, to move scintillator counters from Karlsruhe to
  Tunka after the shutdown of the KASCADE-Grande array in spring 2012. 
  
Scintillation detectors will be installed with a spacing of 150-170\,m and cover an area of about 
1 km$^2$. This will permit to perform a new QUEST-like experiment \cite{cris_06} and to obtain a point of absolute 
energy calibration  at  $3\cdot 10^{16} - 10^{17}$\,eV in 2012-2013.

%

\subsection{HiSCORE station}

The aim of the HiSCORE project \cite{Tlu2009} is to explore the gamma-ray skymap beyond 10
TeV with a future large-area (10-100 km$^2$) wide-angle ($>0$.6 sr) non-imaging cosmic ray and
gamma-ray air shower detector.

HiSCORE is a net of detector stations with 4 PMT-channels equipped with 
Winston cones. The station light-collecting area (0.5 m$^2$) will be a factor
16 larger than that of an Tunka optical detector, resulting in a significantly lower
energy threshold. The first such station will be installed in fall 2011 at the Tunka site.
In 2012-2013, a net of such stations (HiSCORE Engineering Array (EA)) will be installed for 
common operation with Tunka-133. The aim of the HiSCORE EA is to prove methods of EAS 
reconstruction before deployment of a full scale array.


\section{Acknowledgements}  
\label{s6}
This work is supported by Russian Federation Ministery of Science and Education
(G/C 16.518.11.7051, 14.740.11.0890, P681), the Russian Foundation for Basic Research (grants
10-02-00222, 11-02-10005, 11-02-0409, 11-02-91332, 11-02-12138). Authors acknowledge J.Oertlin (KIT) for his 
 help in the analysis of the radio-data.



\end{document}